\documentclass[12pt]{article}

\usepackage{comment}
\usepackage{tikz}
\usepackage{caption}

\usepackage{mathtools}

\usepackage[mathscr]{euscript}

\usepackage{slashed}

\usepackage{amsmath}

\date{}

\title{\textbf{
BRST Analysis and BFV Quantization of the Generalized Quantum Rigid Rotor
}}
\author{ \textbf{Ronaldo Thibes}
\\\\
\textit{\small{Departamento de Ci\^encias Exatas e Naturais}},\\
\textit{\small{Universidade Estadual do Sudoeste da Bahia}},\\
\textit{\small{Rodovia BR 415, km 03, s/n - Itapetinga - BA}},\\
\textit{\small{45700-000 Brazil}}
 }

\usepackage{graphicx}
\usepackage{amssymb,amsfonts,amssymb,amsthm}
\usepackage{amsmath}

\begin{document}

\maketitle

\abstract{We identify a strong similarity among several distinct originally second-class systems, including both mechanical and field theory models, which can be naturally described in a gauge-invariant way.  The canonical structure of such related systems is encoded into a gauge-invariant generalization of the quantum rigid rotor.  We pursue the BRST symmetry analysis and the BFV functional quantization for the mentioned gauge-invariant version of the generalized quantum rigid rotor.  We obtain different equivalent effective actions according to specific gauge-fixing choices, showing explicitly their BRST symmetries.  We apply and exemplify the ideas discussed to two particular models, namely, motion along an elliptical path and the $O(N)$ nonlinear sigma model, showing that our results reproduce and connect previously unrelated known gauge-invariant systems.
}

\section{Introduction}
The study of crafty ingenuous mechanical models has been an important tool for understanding several more involved field and string theory aspects.  Formal concepts with high degree of sophistication such as Grassmann variables, ghost fields, open algebras or BRST symmetry -- common place in quantum field theory, can be realized in particular simple mechanical models, some of them properly designed for field theory comparative purposes
\cite{Plyushchay:1993hs, Loeffelholz:1996cv, Plyushchay:2000hb, Shimizu:2005mq, Shukla:2015wka, Barbosa:2018dmb}.
In this letter, we analyze a generalization of the quantum rigid rotor which can be directly applied to some specific field theory models.  That generalization results in a rich gauge-invariant dynamical system which we examine here in detail at classical and quantum levels.  In particular, we shall discuss its BRST symmetries and pursue its BFV quantization, allowing for an interesting application to the $O(N)$ nonlinear sigma model.

The quantization of constrained dynamical systems has been a challenge since at least the middle of last century from which we may cite the early works \cite{Dirac:1950pj, Anderson:1951ta, Dirac:1951zz, Bergmann:1954tc, Dirac}.
In simple terms, a constrained system is naturally described by a certain set of interdependent dynamical variables.  That means they are not independent variables, but rather must satisfy some specific relations, usually known as constraint equations.  The quantization of such systems can be done along the lines of eliminating some of the dependent variables in terms of the others, or maintaining the constraints at quantum level.  Although the first option may initially seem the most obvious and appropriate one, it is rarely achievable in the proper realistic cases within quantum field theory due to strong technical reasons -- thus we are invariably forced to follow the second path.  Concerning the various possible roads to quantization, passing through operatorial or functional approaches, the quantum description of constrained systems is far from unique.  A key feature regarding the analysis of dynamical systems, in general terms, {\it symmetry} has always played a prominent distinct role.  Knowledge of the symmetry content of specific models has always been more than just helpful, sometimes a decisive means, for understanding dynamical systems -- from simple classical discrete systems up to continuous quantum field theory including string and superstring models.  Actually, when it comes to field theory, most of the models simply cannot be studied without heavily relying on symmetry grounds.  In particular, we may safely say that {\it gauge symmetry} is one of the main cornerstones of the standard model describing nature fundamental interactions, being also present in all current quests for understanding quantum gravity.  Usually, quantization requires the choice of a specific gauge, thus breaking by definition the initial gauge symmetry.  At quantum level, however, a new symmetry replaces the original gauge one -- the {\it BRST symmetry}\footnote{Due to physicists Becchi, Rouet, Stora and Tyutin \cite{Becchi:1974xu, Becchi:1974md, Tyutin:1975qk}.} involving also the ghost fields.  The aim of the present letter is to study the BRST symmetry of a gauge-invariant dynamical system coming from a generalization of the quantum rigid rotor and show its application to specific simple previously unrelated models.

The initial motivation for our work lies on a comparative BRST symmetry study published in 1988 by D.~Nemeschansky, C.~R.~Preitschopf and M.~Weinstein \cite{Nemeschansky:1987xb}.
In that reference, in order to trace analogies to the BRST quantization of Maxwell electrodynamics and Yang-Mills theories, a suitable Lagrangian describing a classical particle moving along a circle is introduced -- with the notations and conventions from \cite{Nemeschansky:1987xb}, we rewrite it here for convenience as
\begin{equation}\label{LNPWclassical}
L=p_r\dot{r}+p_\theta\dot{\theta}-\frac{p_\theta^2}{2mr^2}-\lambda(r-a)\,.
\end{equation}
The authors of \cite{Nemeschansky:1987xb} argue and justify (\ref{LNPWclassical}) on heuristic grounds and then proceed the discussion to obtain a corresponding BRST-invariant version at quantum level, namely  
\begin{equation}\label{LNPW}
L_{NPW}=p_r\dot{r}+p_\theta\dot{\theta}-\frac{1}{2r^2}p_\theta^2-\lambda(r-a)
-b(\dot{\lambda}-p_r)-\frac{1}{2}b^2+\dot{\bar{c}}\dot{c}-\bar{c}c\,,
\end{equation}
again with the same notations, conventions and context from reference \cite{Nemeschansky:1987xb}.
The Lagrangian (\ref{LNPW}) has since then been used as a starting point in a sequel of interesting subsequent papers \cite{Gupta:2009dy, Shukla:2014hea, Shukla:2014spa, Bhanja:2015sha}.

Considering the fact that a single particle moving along a circular path is not, in principle, an intrinsic gauge-invariant system, the above mentioned references call our attention to two important points.  The first one invites us to describe (and compare) that circular motion as a constrained dynamical system.  The circular constraint might be implemented by means of a simple Lagrange multiplier. That leads naturally to a second-class\footnote{We follow the standard nomenclature introduced by Dirac of first- and second-class functions \cite{Dirac:1950pj, Dirac}.  After unraveling the canonical constraint structure through the Dirac-Bergmann algorithm \cite{Dirac:1950pj, Anderson:1951ta}, a phase space function is said to be first-class when its Poisson bracket with each constraint is a linear combination of the constraints set with phase space function coefficients.  Modern reviews of the Dirac-Bergmann algorithm can be seen in \cite{ Hanson:1976cn, Sundermeyer:1982gv, Gitman:1990qh, Henneaux:1992ig, Rothe:2010dzf}.}  constrained system, which we know does not possess gauge invariance.  Noting that the Lagrangian (\ref{LNPWclassical}) somehow does enjoy gauge symmetry, leading to a consistent BRST symmetry in the gauge-fixed version (\ref{LNPW}), brings us to our second point: the fundamental and longstanding constrained systems quantization problem of converting second-class constraints to first-class and producing gauge symmetry.  That is also known as the {\it constraints abelianization} problem.

The usual canonical quantization of second-class systems relies on the replacement of the Poisson brackets by the Dirac ones \cite{Dirac, Hanson:1976cn, Sundermeyer:1982gv, Gitman:1990qh, Henneaux:1992ig, Rothe:2010dzf}.  This easily leads to complications such as nonlocality or operator ordering problems at quantum level.  One way to overcome these and related difficulties comes to converting the second-class constraints to first-class in order to readily produce gauge invariance and take advantage of the BRST symmetry which remains at quantum level even after the gauge-fixing.
The enlargement of the phase space with additional ghost variables permits to use a simpler canonical bracket algebra structure.  Furthermore we have nowadays many powerful functional quantization techniques, such as the Batalin-Fradkin-Vilkovisky (BFV) procedure \cite{Fradkin:1975cq, Batalin:1977pb}, which are properly designed to deal with first-class constraints.  First attempts into the second-class constraints conversion can be seen in \cite{Stueckelberg:1957zz, Wess:1971yu}
where that goal is achieved by the introduction of extra auxiliary variables.  The use of auxiliary variables to promote the abelianization of the second-class constraints has been summarized in the Batalin-Fradkin-Fradkina-Tyutin (BFFT) method 
\cite{Batalin:1986aq, Batalin:1986fm, Batalin:1989dm, Batalin:1991jm},
of which some important applications can be seen in \cite{Amorim:1994np, Amorim:1994ft, Taie:2014oza, Sararu:2016qnx}.  Since then, there has appeared many generalizations of the BFFT ideas, we mention for instance the improved BFFT \cite{Kim:1997psa, Park:1997zj}, the embedding BFFT \cite{Banerjee:1994pp, Oliveira:1998ek, Abreu:2000ip}, the Wotsazek-Neves \cite{Neves:1999jr, Abreu:2000sq} and the gauge-unfixing methods \cite{Anishetty:1992yk, Vytheeswaran:1994np, Neto:2006gt, Neto:2009rm, Monemzadeh:2014wma}.  The very interpretation of some of the second-class constraints as gauge-fixing conditions for other corresponding constraints, which then acquire the status of first-class ones, has also produced an interesting investigative analysis \cite{Harada:1988aj, Mitra:1990mp, Mitra:1990qt}.
In a treatment to constraints abelianization more similar to our spirit here we mention the works \cite{Lyakhovich:2001cm, Batalin:2001hs}
which also do not rely on the introduction of auxiliary variables in phase space.

Aiming at a better understanding and generalization of (\ref{LNPW}) and looking for an alternative general constraints abelianization procedure, we have recently proposed and discussed a few similar models in \cite{Barbosa:2018dmb, Barbosa:2018bng, deOliveira:2019eva}.
Our current idea to be developed below is to start with a generalized Hamiltonian for a system confined to move along a given geometric hypersurface in a gauge-invariant way.  That Hamiltonian can be thought to come from an equivalent second-class system, as shown in reference \cite{deOliveira:2019eva}.  
Hence, gauge-invariance is naturally achieved without the need of auxiliary variables in phase space.
We then proceed to discuss its corresponding BRST symmetries and BFV functional quantization.

Our work is organized as follows.  In section {\bf 2} below, we present the general model, discuss its equations of motion and obtain two possible general gauge-invariant canonical actions. In section {\bf 3}, we introduce the ghost fields algebra with proper Grasssmann variables and construct the conserved BRST charge $\Omega$.  We then show that $\Omega$ consistently generates a BRST symmetry, mixing the original variables with the ghosts and antighosts.
By using the BFV method, we proceed in section {\bf 4} with the functional quantization of the model.  We show that specific functional integrations in the momenta fields produce new effective BRST-invariant actions.  In section {\bf 5}, we exemplify the general results for a particle in motion along an elliptical path and show how (\ref{LNPW}) can be obtained as a particular case.  The $O(N)$ nonlinear sigma model is finally discussed in section {\bf 6} where features previously known as {\it hidden symmetries} are shown to emerge naturally from our general approach.   The application of the ideas developed for the general model are shown to lead to a gauge-invariant action for the $O(N)$ nonlinear sigma model and permit us to obtain a corresponding BRST conserved charge.   We close in section {\bf 7} with our conclusion and final remarks.

\section{The Generalized Quantum Rigid Rotor}
Given a non-degenerated metric function $f^{ij}(q^k)$, a differentiable arbitrary function $T(q^k)$, both depending on the $n$ real variables $q^k$ with $i,j,k = 1,\dots,n$, and an additional real variable $q_0$, we consider the dynamical model defined by the Hamiltonian
\begin{equation}\label{H}
H = W(q^k,p_k) + q_0T(q^k)
\end{equation}
where
\begin{equation}\label{W}
W(q^k,p_k)\equiv\frac{R^{ijkl}T_iT_jp_kp_l}{2f^{ij}T_iT_j}
\end{equation}
with $R^{ijkl}$ denoting the Riemman curvature tensor associated to $f^{ij}$, explicitly given by
\begin{equation}\label{Rijkl}
R^{ijkl}(q^m)\equiv f^{ij}f^{kl} - f^{ik}f^{jl}
\end{equation}
and $T_i$ standing for the $T$ partial derivative with respect to $q^i$, i.e.,
\begin{equation}
T_i\equiv\frac{\partial T}{\partial q^i}
\,.
\end{equation}
The $n$ canonical variables $p_k$ in (\ref{W}) denote the conjugated momenta associated to $q^k$.
The system (\ref{H}) corresponds to a non-trivial generalization of the quantum rigid rotor in $n$ dimensions -- the latter being realized in the very particular case where
$f^{ij}=\delta^{ij}/m$ and
\begin{equation}\label{rr}
T(q^k)=\sqrt{q^i q^i}-a
\,.
\end{equation}
In the rigid rotor,
the constant real number $a$ in equation (\ref{rr}) represents the radius of a ($n-1$)-sphere in $\mathbb{R}^n$ where a mass $m$ particle is confined to move.  In the general case, the system (\ref{H}) naturally evolves along the geometric hypersurface $T(q^k)=0$, which can be chosen at best convenience.

The Hamilton equations corresponding to the general dynamical system (\ref{H}) can be directly computed and written out as
\begin{equation}\label{Heqs}
\begin{cases}
{\dot{q}}^i=
\displaystyle\frac{R^{ijkl}T_kT_lp_j}{f^{ij}T_iT_j}\,,\\
{\dot{p}}_i=-W_i-q_0T_i \,,\\
T=0\,,
\end{cases}
\end{equation}
where
\begin{equation}\label{Wi}
W_i\equiv\frac{\partial W}{\partial q^i}=
\frac{
\left[
\left(R^{mjkl}_{\,\,\,\,\,\,\,\,\,\,\,\,,i}f^{rs}-R^{mjkl}f^{rs}_{\,\,\,\,\,\,,i}\right)T_s+2f^{rk}R^{msjl}T_{si}
\right]T_rT_mT_jp_kp_l
}
{2f^{ij}f^{kl}T_iT_jT_kT_l}
\,,
\end{equation}
with $T_{ij}$ denoting the second-order derivative
\begin{equation}
T_{ij}
\equiv\frac{\partial^2 T}{\partial q^j \partial q^i}
\,.
\end{equation}
The indexes $l,m,r,s$ in equation (\ref{Wi}) also run from $1$ to $n$.
Obtaining a classical solution for this system consists in solving for the $n$ coordinates $q^i$, as well as $q_0$, as functions of the time evolution parameter $t$.  We note however that the Hamiltonian (\ref{H}) defines a {\it constrained} system with $n-1$ degrees of freedom.
In fact, by introducing the compact notation
\begin{equation}\label{wB}
w\equiv f^{ij}T_iT_j\,,\,\,\,\,\,\,\,\,B\equiv f^{ij}T_ip_j\,,
\,\,\mbox{and}\,\,{\cal O}_i^{\,\,\,j}\equiv \delta_i^j-\frac{T_if^{jk}T_k}{w}\,,
\end{equation}
we may easily determine $q_0$ in (\ref{Heqs}) as
\begin{equation}
q_0 = -\frac{f^{ij}T_i(\dot{p}_j+W_j)}{w}
\end{equation}
and rewrite the remaining $2n$ first-order differential equations as
\begin{equation}\label{rde}
\begin{cases}
\displaystyle\dot{q}^i=f^{ij}\left(p_j-\frac{B}{w}T_j\right)\,,\\
{\cal O}_i^{\,\,\,j}(\dot{p}_j+W_j)=0\,.
\end{cases}
\end{equation}
As the operator ${\cal O}_i^{\,\,\,j}$ is not invertible, due to the zero-mode ${\cal O}_i^{\,\,\,j}T_j=0$, the momenta $p_i$ cannot be 
univocally determined.  Furthermore, the velocities $\dot{q}^i$ are also interrelated by $T_i\dot{q}^i=0$ as immediately follows from (\ref{rde}), or equivalently from the last equation in (\ref{Heqs}) implying $\dot{T}=T_i\dot{q}^i=0$.

To be more precise, the Hamiltonian (\ref{H}) characterizes a {\it gauge-invariant} system.  This can be seen by introducing the momentum variable $p_0$ conjugated to $q_0$, a Lagrange multiplier $\lambda$, and considering the corresponding canonical action given by
\begin{equation}\label{Sc}
S_c=\int_{t_1}^{t_2}dt\,\left[
\dot{q}^ip_i+\dot{q}_0p_0-W-q_0T-\lambda p_0\right]
\,,
\end{equation}
with $W$ given by (\ref{W}).
Then it is clear by inspection that if $\epsilon$ and $\epsilon_0$ denote arbitrary time-dependent parameters, the canonical action (\ref{Sc}) is left 
invariant under the gauge transformations
\begin{equation}
\begin{cases}\label{ginv}
\delta q_0 = \epsilon_0+\dot{\epsilon}\,,\\
\delta \lambda = \dot{\epsilon}_0+\ddot{\epsilon}\,,\\
\delta p_i = \displaystyle\frac{\epsilon_0TT_i}{\dot{T}}-\epsilon T_i\,,
\end{cases}
\end{equation}
generated by the two first-class constraints $p_0$ and $T$.  The variables $q_0$ and  $\lambda$ act as two Lagrange multipliers enforcing the constraints $T=0$ and $p_0=0$.  Actually, since besides in the term $\lambda p_0$ in (\ref{Sc}), the momentum $p_0$ appears only in the kinetic part of the action, we may dispense with it and its Lagrange multiplier partner altogether from the action and, alternatively, consider the reduced more compact simpler action
\begin{equation}\label{S}
S=\int_{t_1}^{t_2}dt\,\left[
\dot{q}^ip_i-H\right]
\,,
\end{equation}
with Hamiltonian function $H$ given by equation (\ref{H}).  The equivalent alternative action (\ref{S}) is also gauge-invariant under the variations
\begin{equation}
\begin{cases}\label{ginv}
\delta q_0 = \epsilon_0+\dot{\epsilon}\,,\\
\delta p_i = \displaystyle\frac{\epsilon_0TT_i}{\dot{T}}-\epsilon T_i\,,
\end{cases}
\end{equation}
for arbitrary gauge parameters $\epsilon_0$ and $\epsilon$.
The main purpose of the present paper is to describe the BRST symmetries associated to the canonical gauge-invariant action (\ref{Sc}), discuss its quantization and clearly show in the examples how it can be used to produce gauge invariance for originally second-class specific models.  This will be done in the next sections.

\section{BRST Symmetries}
The Becchi-Rouet-Stora-Tyutin (BRST) symmetry has first appeared in the groundbreaking works \cite{Becchi:1974xu, Becchi:1974md, Tyutin:1975qk}.  Since then, it has been around almost ubiquitously along the quantization process of gauge-invariant field theories.  Surviving after gauge-fixing, the BRST symmetry involves the ghost fields and plays a fundamental role in the renormalization program of quantum field theory \cite{Piguet:1995er}.  Besides the original references \cite{Becchi:1974xu, Becchi:1974md, Tyutin:1975qk}, the BRST quantization scheme can be found in many reviews and textbooks as for instance
\cite{Nemeschansky:1987xb, Henneaux:1992ig, Henneaux:1985kr, Niemi:1988bf, Becchi:1996yh, Hong:2015fvc}.
In order to study the BRST symmetries associated to the model (\ref{H}) at quantum level, we turn into the extended phase space by introducing the Grassmann variables pairs $(C, {\cal P})$ with ghost number one and $({\bar C}, \bar{\cal P})$ with ghost number minus one.  This permits  the construction of the BRST charge from the constraints $T$ and $p_0$ as
\begin{equation}\label{Omega}
\Omega=CT-ip_0{\cal P}\,,
\end{equation}
possessing odd parity and ghost number one.  The complete phase space variables set, displayed in the table below,
satisfies the following canonical relations
\begin{equation}
\left[ q^i,p_j \right] = \delta_j^i\,,\,\,\,\,
\left[ q_0,p_0 \right] = 1\,,
\end{equation}
\begin{equation}
\left[ C, \bar{\cal P} \right] =
\left[ \bar{C}, {\cal P} \right] = -1
\,.
\end{equation}
Note
\captionsetup{labelformat=empty}
\begin{table}[ht]
\centering
{\small\begin{tabular}{lrrrrrrrr} \hline\hline&\rule{0pt}{3ex}$q^i$&$p_i$&$q_0$&$p_0$&$C$&$\bar{C}$&${\cal P}$&${\bar{\cal P}}$\\
\hline
Grassmann parity&$0$&$0$&$0$&$0$&$1$&$1$&$1$&$1$\\
ghost number&$0$&$\phantom{-}0$&$\phantom{-}0$&$\phantom{-}0$&$\phantom{-}1$&$-1$&$\phantom{-}1$&$-1$
\\
\hline\hline
\end{tabular}}
\caption{Grassmann parity and ghost numbers} \label{t1}
\end{table}
in particular that besides the original phase space variables and the proper BRST prescribed ghost variables associated to the constraints $T$  and $p_0$, no additional auxiliary variables need to be introduced.

Within this framework, we can check that the constraints set $(T,p_0)$ is indeed first-class and forms a closed algebra with the Hamiltonian (\ref{H}) given by
\begin{equation}
\left[T,H\right]=\left[T,p_0\right]=0\,,
\end{equation}
\begin{equation}
\left[p_0,H\right]=-T\,.
\end{equation}
Moreover, we define the BRST transformations generated by $\Omega$ 
for an arbitrary extended phase space function $F$ as
\begin{equation}\label{sF}
sF\equiv \left[ F,\Omega\right]\,.
\end{equation}
Concerning the canonical variables basis, this leads to the non-null explicit variations
\begin{equation}\label{s}
sp_i=-CT_i\,,\,\,\,\,\,sq_0=-i{\cal P}\,,\,\,\,\,\,s\bar{C}=ip_0\,,\,\,\,\,\,s\bar{\cal P}=-T\,.
\end{equation}
The BRST operator $s$ can be easily checked to be nillpotent and have ghost number and parity one, as a direct consequence from its definition (\ref{sF}).
Next, following the original references
\cite{Becchi:1974xu, Becchi:1974md, Tyutin:1975qk, Henneaux:1985kr}, we construct the extended BRST-invariant Hamiltonian given by
\begin{equation}
H_{ext} = W - \left[ \Psi,\Omega \right]
\end{equation}
where $\Psi$ denotes an arbitrary gauge-fixing fermion.

For definiteness, we work with the usual standard form
\begin{equation}
\Psi = i\bar{C}\chi + q_0\bar{\cal P}
\end{equation}
for the gauge-fixing fermion and consider for the open function $\chi(q^i,p_i,q_0,p_0)$ the two following
particular possibilities:
\begin{equation}\label{chi1}
\chi_1=\dot{q}_0+\frac{B^3}{w{\cal T}^2}-\frac{p_0B^4}{2w{\cal T}^4}
\end{equation}
and
\begin{equation}\label{chi2}
\chi_2=\frac{B^3}{3w{\cal T}^2}
\,,
\end{equation}
where we have defined
\begin{equation}\label{calT}
{\cal T}=T(q^i,p_i)-T(0,0)
\,.
\end{equation}
The first one (\ref{chi1}) depends on $\dot{q}_0$ and leads to the velocity-dependent BRST-invariant Hamiltonian
\begin{equation}\label{Hext1}
H_{ext{\it 1}}=W+q_0T+\dot{q}_0p_0-i{\cal P}\bar{\cal P}
-{\bar C}\dot{\cal P}+i\bar{C}\left(3-2\frac{p_0B}{{\cal T}^2}\right)\frac{B^2C}{{\cal T}^2}
+\frac{p_0B^3}{w{\cal T}^2}
\left(1-\frac{p_0B}{2{\cal T}^2}\right)
\end{equation}
while the second one (\ref{chi2}) leads to
\begin{equation}\label{Hext2}
H_{ext{\it 2}}=W+q_0T+\frac{p_0B^3}{3w{\cal T}^2}
+i\bar{C}\frac{B^2}{{\cal T}^2}C-i{\cal P}\bar{\cal P}
\,.
\end{equation}
Note that both extended Hamiltonians (\ref{Hext1}) and (\ref{Hext2}) are invariant under the BRST transformations (\ref{s}).
Interesting enough, after functional integration on the ghost momenta through the BFV quantization scheme, the gauge-fixing (\ref{chi1}) leads to an effective action independent from ghosts time-derivative whilst the opposite happens to (\ref{chi2}).  This will be explicitly shown in the next section where we pursue the system functional quantization.

\section{Functional Quantization}
In this section we consider the Batalin-Fradkin-Vilkovisky (BFV) functional quantization of the model (\ref{H}).
The BFV quantization scheme 
\cite{Fradkin:1975cq, Batalin:1977pb, Henneaux:1985kr}
takes place in the extended phase space including the ghosts being well suited for Hamiltonian systems with first-class constraints, as is the case for our current working model.
We start by introducing the Green functions generating functional given by
\begin{equation}\label{Z}
Z = \int [d\varpi] \exp \left[ \frac{i}{\hbar} S_{ext} \right]
\end{equation}
where $[d\varpi]$ stands for the Liouville functional integration measure including all phase space variables, i.e.,
\begin{equation}
[d\varpi] = [dq^i][dp_i][dq_0][dp_0][dC][d\bar{C}][d{\cal P}][d\bar{\cal P}]
\end{equation}
and
\begin{equation}\label{Sext}
S_{ext} = \int_{t_1}^{t_2}\,dt\,
\left(
\dot{q}^i p_i +\dot{q}_0 p_0 +\dot{C}\bar{\cal P}
+\dot{\cal P}\bar{C}-H_{ext}
\right)
\end{equation}
denotes the extended action constructed from the BRST-invariant Hamiltonian which we obtained in the last section.  

Considering first the velocity-dependent gauge-fixing (\ref{chi1}), substituting $H_{ext}=H_{ext{\it 1}}$ from (\ref{Hext1}) in the extended action (\ref{Sext}) and performing the functional integration in equation (\ref{Z}) in the variables $p_0$, $\cal P$ and $\bar{\cal P}$ we may rewrite the generating functional as
\begin{equation}
Z=\int [d\mu]
\exp \left[ \frac{i}{\hbar} S_{eff{\it 1}} \right]
\end{equation}
with the reduced integration measure
\begin{equation}
[d\mu]=[dq^i][dp_i][dq_0][dC][d\bar{C}]
\end{equation}
and effective action
\begin{equation}\label{Seff1}
S_{eff{\it 1}}=\int_{t_1}^{t_2}\,dt\left[
\dot{q}^ip_i-\frac{1}{2}f^{ij}p_ip_j-q_0T-i\bar{C}\frac{B^2}{{\cal T}^2}C
\right]
\,.
\end{equation}
The neat extended action (\ref{Seff1}) is invariant under a BRST transformation involving its solely own variables given by
\begin{equation}\label{BRST1}
\delta_1 p_i = -CT_i\,,\,\,\,\,\delta_1 q_0 = \dot{C}\,,\,\,\,\,
\delta_1 \bar{C}=i\frac{{\cal T}^2}{B}+\bar{C}\frac{w}{B}C\,,\,\,\,\,
\delta_1 B = -wC\,,
\end{equation}
as well as under a corresponding anti-BRST transformation
\begin{equation}\label{aBRST1}
\bar{\delta}_1 p_i = \bar{C}T_i\,,\,\,\,\,\bar{\delta}_1 q_0 = -\dot{\bar{C}}\,,\,\,\,\,
\bar{\delta}_1 {C}=i\frac{{\cal T}^2}{B}+\bar{C}\frac{w}{B}C\,,\,\,\,\,
\bar{\delta}_1 B = w\bar{C}\,,
\end{equation}
which exchanges the roles between the ghosts $C$ and $\bar{C}$.
Note that both transformations (\ref{BRST1}) and (\ref{aBRST1}) are nillpotent.

As a second option, we consider next the alternative gauge-fixing (\ref{chi2}) and its corresponding extended Hamiltonian (\ref{Hext2}) which, inserted into (\ref{Sext}) and (\ref{Z}), leads, after functional integration in the ghost momenta $\cal P$ and $\bar{\cal P}$, to
\begin{equation}
Z=\int [d\mu']
\exp \left[ \frac{i}{\hbar} S_{eff{\it 2}} \right]
\end{equation}
with
\begin{equation}
[d\mu']=[dq^i][dp_i][dq_0][dp_0][dC][d\bar{C}]
\end{equation}
and
\begin{equation}\label{Seff2}
S_{eff{\it 2}}=\int_{t_1}^{t_2}\,dt\left[
\dot{q}^ip_i-W-q_0T
+p_0\left(\dot{q}_0-\frac{B^3}{3w{\cal T}^2}\right)
-i\bar{C}\left(\frac{B^2}{{\cal T}^2}+\frac{d^2}{dt^2}\right)C
\right]
\,.
\end{equation}
Similarly to (\ref{Seff1}), the effective action (\ref{Seff2}) also enjoys invariance under nillpotent BRST and anti-BRST transformations mixing its own variables, namely,
\begin{equation}
\delta_2 p_i = -CT_i\,,\,\,\,\,\delta_2 q_0 = \dot{C}\,,\,\,\,\,
\delta_2\bar{C}=ip_0\,,\,\,\,\,\delta_2 B = -wC
\,,
\end{equation}
and
\begin{equation}
\bar{\delta}_2 p_i = \bar{C}T_i\,,\,\,\,\,\bar{\delta}_2 q_0 = -\dot{\bar{C}}\,,\,\,\,\,
\bar{\delta}_2{C}= ip_0  \,,\,\,\,\,\bar{\delta}_2 B = w\bar{C}
\,.
\end{equation}
 We have achieved our goal concerning the BFV quantization of the gauge-invariant dynamical system (\ref{H}) and exploring its BRST symmetries at quantum level.  In the next sections we shall discuss two specific examples.

\section{Mechanical Example: Elliptical Path}
As a first illustrative example of the ideas previously discussed, in the current section we consider a mass $m$ particle in two dimensions confined to move along an elliptical path given by the Cartesian equation
\begin{equation}\label{ellipse}
\frac{x^2}{a^2}+\frac{y^2}{b^2}=1
\end{equation}
where $x$ and $y$ denote Cartesian coordinates and $a$ and $b$ the two ellipse major and minor axis.
As mentioned in the Introduction, a BRST-invariant action describing the motion of a particle along a circular path has appeared in \cite{Nemeschansky:1987xb} and has been the starting point for a handful of further theoretical developments from which we mention 
\cite{Gupta:2009dy, Shukla:2014hea, Shukla:2014spa}.
It will be interesting to compare our results with those of references \cite{Nemeschansky:1987xb, Gupta:2009dy, Shukla:2014hea, Shukla:2014spa} for the particular case $b=a$ in which the ellipse (\ref{ellipse}) degenerates to a circle.  For this reason we choose the form of the main constraint equation as
\begin{equation}
T(x,y)=\sqrt{ab}\left[
\sqrt{(x/a)^2+(y/b)^2}-1
\right]
\end{equation}
and,
corresponding to equation (\ref{H}), define the Hamiltonian function as
\begin{equation}\label{Hellipse}
H(x,y,z,p_x,p_y)=\frac{(xp_yb/a-yp_xa/b)^2}{2m(x^2b^2/a^2+y^2a^2/b^2)}
+z\sqrt{ab}\left[
\sqrt{(x/a)^2+(y/b)^2}-1
\right]
\,.
\end{equation}
The resulting model (\ref{Hellipse}) was also briefly discussed in \cite{deOliveira:2019eva}  without mention to BRST symmetry.
For notational convenience, we
write down further
\begin{equation}
{\cal T}\equiv T+\sqrt{ab} = \sqrt{x^2b/a+y^2a/b}
\end{equation}
in accordance to (\ref{calT}),
and note that the general definitions (\ref{wB}) particularize here to
\begin{equation}
w=\frac{x^2b^2/a^2+y^2a^2/b^2}{m{\cal T}^2}
\end{equation}
and
\begin{equation}
B=\frac{xp_xb/a+yp_ya/b}{m{\cal T}}
\,.
\end{equation}

Proceeding next to the extended phase space, we introduce the canonical momentum $p_z$ and ghost variables $(C,{\bar C},{\cal P},{\bar{\cal P}})$ in order to obtain a BRST-invariant Hamiltonian.  By choosing a gauge-fixing of the form (\ref{chi2}) we may write a concise no-time-derivatives extended Hamiltonian
\begin{eqnarray}\label{Hext2ellipse}
H_{ext{\it 2}}&=&H+
\frac{p_z(xp_xb/a+yp_ya/b)^3)}
{3m^2(x^2b^2/a^2+y^2a^2/b^2)(x^2b/a+y^2a/b)^{2/3}}
\nonumber\\
&&+\,
i{\bar{C}}\frac{(xp_xb/a+yp_ya/b)^2}{m^2(x^2b/a+y^2a/b)}C
+i\bar{\cal P}{\cal P}
\,,
\end{eqnarray}
with $H$ given by equation (\ref{Hellipse}) and
which can be checked to be invariant under the nillpotent BRST transformations
\begin{equation}
sx=sy=0\,,\,\,\,sz=-i{\cal P}\,,
\end{equation}
\begin{equation}
sp_x = -\frac{xbC}{a\sqrt{x^2b/a+y^2a/b}} \,,\,\,\,sp_y= -\frac{yaC}{b\sqrt{x^2b/a+y^2a/b}}\,,\,\,\,sp_z=0\,,
\end{equation}
\begin{equation}
sC=s{\cal P}=0\,,\,\,\,s\bar{C}=ip_z\,,\,\,\,
s\bar{\cal P}=-\sqrt{ab}\left[
\sqrt{(x/a)^2+(y/b)^2}-1
\right]
\,.
\end{equation}
In the particular case $b=a$, by using polar coordinates\footnote{From $x=r\cos\theta$ and $y=r\sin\theta$ we have $p_r=\cos\theta\,p_x+\sin\theta\,p_y$ and $p_\theta=-r\sin\theta p_x +r\cos\theta p_y$.
} $(r,\theta,p_r,p_\theta)$, the extended Hamiltonian (\ref{Hext2ellipse}) may be rewritten as
\begin{equation}
H_{ext{\it 2}}=\frac{p_\theta^2}{2mr^2}+z(r-a)+
\frac{p_zp_r^3}{3m^2r^2}+i\bar{C}\frac{p_r^2}{m^2r}C+
\bar{\cal P}{\cal P}
\end{equation}
being invariant under the non-null BRST transformations
\begin{equation}
sz=-i{\cal P}\,,\,\,\,sp_r=-C\,,\,\,\,
s\bar{C}=ip_z\,,\,\,\,s{\bar{\cal P}}=-(r-a)\,.
\end{equation}

As shown last section for the general case, after obtaining the generating functional via the BFV quantization scheme, it is possible to produce a simpler effective action.  For instance corresponding to the equation (\ref{Seff1}) we have the effective action
\begin{eqnarray}\label{Seff1ellipse}
S_{eff{\it 1}}&=&\int_{t_1}^{t_2}\,dt\left[
\dot{x}p_x+\dot{y}p_y-\frac{p_x^2+p_y^2}{2m}
-z\sqrt{ab}\left[
\sqrt{(x/a)^2+(y/b)^2}-1
\right]
\right.\nonumber\\&&\left.
-i\bar{C}
\frac{(xp_xb/a+yp_ya/b)^2}{m^2(x^2b/a+y^2a/b)}
C
\right]\,,
\end{eqnarray}
while a version depending on the ghosts time derivatives can be obtained from (\ref{Seff2}) as
\begin{eqnarray}\label{Seff2ellipse}
S_{eff{\it 2}}&\!=&\!\!\int_{t_1}^{t_2}\,dt\left[
\dot{x}p_x+\dot{y}p_y
-\frac{(xp_yb/a-yp_xa/b)^2}{2m(x^2b^2/a^2+y^2a^2/b^2)}
-z\sqrt{ab}\left[
\sqrt{(x/a)^2+(y/b)^2}-1
\right]
\right.\nonumber\\&
+\,p_z&\left.\!\!\!\!\!\!\left(
\dot{z}-\frac{(xp_xb/a+yp_ya/b)^3}{3m^2{\cal T}^3(x^2b^2/a^2+y^2a^2/b^2)}\right)
+i\bar{C}\left(
\frac{(xp_xb/a+yp_ya/b)^2}{m^2{\cal T}^4}+\frac{d^2}{dt^2}\right)C
\right]\,.
\end{eqnarray}
In polar coordinates, for the particular case $b=a$, the two last effective actions read respectively
\begin{equation}
S_{eff{\it 1}}=\int_{t_1}^{t_2}\,dt\left[
\dot{r}p_r+\dot{\theta}p_\theta-\frac{p_r^2}{2m}-\frac{p_\theta^2}{2mr^2}
-z(r-a)-i\bar{C}\frac{p_r^2}{m^2r^2}C
\right]
\end{equation}
and
\begin{eqnarray}
S_{eff{\it 2}}&=&\int_{t_1}^{t_2}\,dt\left[
\dot{r}p_r+\dot{\theta}p_\theta-\frac{p_\theta^2}{2mr^2}
-z(r-a)
\nonumber\right.\\&&\left.
+p_z\left(\dot{z}-\frac{p_r^3}{3m^2r^2}\right)
-i\bar{C}\left(\frac{p_r^2}{m^2r^2}+\frac{d^2}{dt^2}\right)C
\right]\,,
\end{eqnarray}
being invariant under the following BRST and anti-BRST transformations:
\begin{equation}\label{BRST1circle}
\delta_1 p_r = -C\,,\,\,\,\,\delta_1 z = \dot{C}\,,\,\,\,\,
\delta_1 \bar{C}=i\frac{mr^2}{p_r}+\frac{\bar{C}C}{p_r}\,,
\end{equation}
\begin{equation}\label{aBRST1circle}
\bar{\delta}_1 p_r = \bar{C}\,,\,\,\,\,\bar{\delta}_1 z = -\dot{\bar{C}}\,,\,\,\,\,
\bar{\delta}_1 {C}=i\frac{mr^2}{p_r}+\frac{\bar{C}C}{p_r}\,,
\end{equation}
\begin{equation}
\delta_2 p_r = -C\,,\,\,\,\,\delta_2 z = \dot{C}\,,\,\,\,\,
\delta_2\bar{C}=ip_z\,,
\end{equation}
and
\begin{equation}
\bar{\delta}_2 p_r = \bar{C}\,,\,\,\,\,\bar{\delta}_2 z = -\dot{\bar{C}}\,,\,\,\,\,
\bar{\delta}_2{C}= ip_z
\,.
\end{equation}

\section{Field Theory Example:  The O(N) Nonlinear Sigma Model}
The O(N) nonlinear sigma model can be described by the Lagrangian density
\begin{equation}\label{ONL}
{\cal L} = \frac{1}{2f}\partial_\mu\phi^a\partial^\mu\phi^a-\frac{\varphi}{2}(\phi^a\phi^a-F^2)
\end{equation}
where $\phi^a$ denotes a multiplet of real scalar fields with $a=1,\dots,N$ and $\varphi$ stands for an additional scalar field. 
We consider the model defined in Minkowsky space, with $D-1$ spatial dimensions and metric convention diag $\eta^{\mu\nu}=(1,-1,\dots,-1)$.  Furthermore, in this section we use natural units $\hbar=c=1$ and standard relativistic notation -- Greek indexes run through $\mu,\nu=0,\dots,D-1$ and middle alphabet Latin indexes $i,j,=1\dots,D-1$.
The two constant real parameters $f$ and $F$ in (\ref{ONL})  provide room for dimensional balance, according to the desired specific application and effective number $D$ of space-time dimensions.  In principle we could say that the Lagrangian density (\ref{ONL}) promotes an interaction between the two scalar fields $\phi^a$ and $\varphi$ through the cubic term
\begin{equation}
{\cal L}_{int} = -\frac{1}{2}\varphi\phi^a\phi^a
\,,
\end{equation}
however, we see that actually $\varphi$ plays the role of a Lagrangian multiplier field, enforcing the constraint\footnote{Of course the factor $1/2$ in the constraint equation is just for convenience.}
\begin{equation}\label{T}
T=\frac{1}{2}(\phi^a\phi^a-F^2)
\end{equation}
which fixes the norm of $\phi^a$.  Therefore the field $\phi^a$ takes value along an $(N-1)$-sphere $S^{N-1}$ and the Lagrangian density (\ref{ONL}) is symmetric with respect to $O(N)$ transformations $\phi^a\rightarrow T^{ab}\phi^b$ with $T^{ab}\in O(N)$.
Besides (\ref{T}), the nonlinear field equations for the Lagrangian density (\ref{ONL}) read
\begin{equation}
(\Box+f\varphi)\phi^a=0\,.
\end{equation}
The canonical structure of this model has been analyzed for instance in references
\cite{Maharana:1983cs, Hong:2002vr}.  Following the standard Dirac-Bergmann algorithm,
we obtain four second-class constraints which we define and write here as
\begin{equation}
\begin{cases}
\Xi_1 = \pi_\varphi\,,
\\
\Xi_2 \equiv T =\frac{1}{2}(\phi^a\phi^a-F^2)\,,
\\
\Xi_3 = \phi^a\pi^a\,,
\\
\Xi_4 = f^2\pi^a\pi^a-f\varphi\phi^a\phi^a+\phi^a\partial_i\partial_i\phi^a\,,
\end{cases}
\end{equation}
where $\pi^a$ and $\pi_\phi$ denote the momenta fields canonically conjugated respectively to $\phi^a$ and $\varphi$.
Actually, the first constraint $\Xi_1=\pi_\varphi$ above is somewhat artificial simply signaling that the field $\varphi$ has no dynamics, acting as a mere Lagrange multiplier.  In this sense it is possible to consider a smaller set of constraints if one follows, for instance, the symplectic Faddeev-Jackiw-BarcelosNeto-Wotzasek (FJBW) iterative approach
\cite{Faddeev:1988qp, BarcelosNeto:1991kw}
as can be seen in references \cite{Hong:2002vr, Foussats:1997aa}.

Following the main course of section {\bf 2}, the $O(N)$ nonlinear sigma model (\ref{ONL}) can be easily cast into a gauge-invariant form.  In fact, equations (\ref{wB}) particularize now to
\begin{equation}\label{wBONL}
w\equiv f\phi^a\phi^a\,,\,\,\,\,\,\,\,\,B\equiv f\phi^a\pi^a\,,
\,\,\,\,\,\mbox{ and }\,\,\,\,\,{\cal O}^{ab}\equiv \delta^{ab}-\frac{\phi^a\phi^b}{\phi^c\phi^c}\,,
\end{equation}
and corresponding to (\ref{Sc}) we have the $O(N)$ nonlinear sigma model canonical action
\begin{eqnarray}\label{ScONL}
S_c&=&\int d^Dx \left\{ \dot{\phi}^a\pi^a+\dot{\varphi}\pi_\varphi
-\frac{f}{2}\left(
\pi^a\pi^a-\frac{\phi^a\phi^b\pi^a\pi^b}{\phi^c\phi^c}
\right)
-\frac{1}{2f}\partial_i\phi^a\partial_i\phi^a
\right.\nonumber\\&&\left.
-\frac{\varphi}{2}\left(\phi^a\phi^a-F^2\right)
-\theta\pi_\varphi
\phantom{\left(\frac{\phi^a\phi^b\pi^a\pi^b}{\phi^c\phi^c}
\right)}\!\!\!\!\!\!\!\!\!\!\!\!\!\!\!\!\!\!\!\!\!\!\!\!\!\!\!\!\!\!\!\!\!\!
\right\}
\,,
\end{eqnarray}
which can be checked to be gauge-invariant under
\begin{equation}
\begin{cases}\label{ginvONL}
\delta \varphi = \epsilon_0+\dot{\epsilon}\,,\\
\delta \theta = \dot{\epsilon}_0+\ddot{\epsilon}\,,\\
\delta \pi^a = \displaystyle\frac{\epsilon_0\left(\phi^b\phi^b-F^2\right)\phi^a}{2\phi^c\dot{\phi^c}}-\epsilon \phi^a\,,
\end{cases}
\end{equation}
for arbitrary space-time dependent gauge parameters $\epsilon_0$ and $\epsilon$.
In equations (\ref{ScONL}) and (\ref{ginvONL}), $\theta$ denotes a Lagrange multiplier field for the constraint $\pi_\varphi$.
Or else, for those who do not like too many constraints and unnecessary redundant variables, corresponding to (\ref{S})
we have the more compact action
\begin{equation}\label{SONL}
S=\int d^Dx \left\{ \dot{\phi}^a\pi^a-
\frac{f}{2}\left(
\pi^a\pi^a-\frac{\phi^a\phi^b\pi^a\pi^b}{\phi^c\phi^c}
\right)
-\frac{1}{2f}\partial_i\phi^a\partial_i\phi^a
-\frac{\varphi}{2}\left(\phi^a\phi^a-F^2\right)
\right\}
\,,
\end{equation}
also gauge-invariant under
\begin{equation}
\delta \varphi = \epsilon_0+\dot{\epsilon}\,\,,\,\,\,\,\,\,
\delta \pi^a = \displaystyle\frac{\epsilon_0\left(\phi^b\phi^b-F^2\right)\phi^a}{2\phi^c\dot{\phi^c}}-\epsilon \phi^a\,.
\end{equation}

Based on the general discussion carried out in section {\bf 2}, we have been able to exhibit a gauge-invariant description for the $O(N)$ nonlinear sigma model.  Henceforth we proceed with the quantization process.
Enlarging the phase space including the appropriate Grassmann variables $(C,\bar{C},{\cal P},\bar{\cal P})$ we obtain, from equation (\ref{Omega}) the BRST charge
\begin{equation}
\Omega=\int d^{D-1}x\left[
\frac{1}{2}C\left(\phi^a\phi^a-F^2\right)-i\pi_\varphi{\cal P}
\right]\,,
\end{equation}
which generates the BRST transformations corresponding to (\ref{s}) given here explicitly by
\begin{equation}\label{sONL1}
s\pi^a = -C\phi^a\,,\,\,\,\,\,\,s\varphi=-i{\cal P}\,,
\end{equation}
\begin{equation}\label{sONL2}
s\bar{C}=i\pi_\varphi\,,\,\,\,\,\,\,s{\bar{\cal P}}=-\frac{1}{2}\left(\phi^a\phi^a-F^2\right)\,.
\end{equation}
A gauge-fixed BRST-invariant Hamiltonian can then be readily constructed, for instance from (\ref{Hext2}) using relations (\ref{wBONL}) and
\begin{equation}
{\cal T}\equiv{\phi^a\phi^a}\,,
\end{equation}
as
\begin{eqnarray}\label{Hext2ONL}
H_{ext2}&=&\int d^{D-1}x
\left[
\frac{f}{2}\left(
\pi^a\pi^a-\frac{\phi^a\phi^b\pi^a\pi^b}{\phi^c\phi^c}
\right)
+\frac{1}{2f}\partial_i\phi^a\partial_i\phi^a
+\frac{\varphi}{2}\left(\phi^a\phi^a-F^2\right)
\right.\nonumber\\&&\left.
+\frac{f^2\pi_\varphi\phi^a\phi^b\phi^c\pi^a\pi^b\pi^c}
{3\phi^d\phi^e\phi^f\phi^d\phi^e\phi^f}
+f^2i\bar{C}\frac{\phi^a\phi^b\pi^a\pi^b}
{\phi^c\phi^d\phi^c\phi^d}C
+i\bar{\cal P}{\cal P}
\right]
\,.
\end{eqnarray}
A straightforward calculation shows that (\ref{Hext2ONL}) is indeed invariant under the BRST transformations (\ref{sONL1}) and (\ref{sONL2}).
Following the steps from section {\bf 4}, c.f. equations (\ref{Seff1}) and (\ref{Seff2}), we may also write down the effective actions
\begin{eqnarray}
S_{eff\it 1} &=& \int d^Dx \left\{ \dot{\phi}^a\pi^a
-\frac{f}{2}\pi^a\pi^a
-\frac{1}{2f}\partial_i\phi^a\partial_i\phi^a
\right.\nonumber\\&&\left.
-\frac{\varphi}{2}\left(\phi^a\phi^a-F^2\right)
-if^2\bar{C}\left[\frac{\phi^a\phi^b\pi^a\pi^b}{\phi^c\phi^c\phi^d\phi^d}\right]C
\right\}
\end{eqnarray}
and
\begin{eqnarray}
S_{eff\it 2} &=& \int d^Dx \left\{ \dot{\phi}^a\pi^a
+\pi_\varphi\dot{\varphi}
-\frac{f}{2}\pi^a\pi^a
+f\frac{\phi^a\phi^b\pi^a\pi^b}{2\phi^c\phi^c}
-\frac{1}{2f}\partial_i\phi^a\partial_i\phi^a
\right.\nonumber\\&&\left.
-\frac{\varphi}{2}\left(\phi^a\phi^a-F^2\right)
-\frac{f^2\pi_\varphi\phi^a\phi^b\phi^c\pi^a\pi^b\pi^c}
{3\phi^d\phi^e\phi^f\phi^d\phi^e\phi^f}
-if^2\bar{C}\left[\frac{\phi^a\phi^b\pi^a\pi^b}{\phi^c\phi^c\phi^d\phi^d}\right]C
-i\bar{C}\ddot{C}
\right\}
\end{eqnarray}
which can be checked to be invariant under the nillpotent BRST
\begin{equation}
\delta_1\pi^a=-C\phi^a\,,\,\,\,\,\,\delta_1\varphi=\dot{C}\,,\,\,\,\,\,\delta_1\bar{C}=i\frac{\phi^a\phi^b\phi^a\phi^b}{f\phi^c\pi^c}+\bar{C}\frac{\phi^a\phi^b}{\phi^c\phi^c}C
\,,
\end{equation}
\begin{equation}
\delta_2\pi^a=-C\phi^a\,,\,\,\,\,\,\delta_2\varphi=\dot{C}\,,\,\,\,\,\,\delta_2\bar{C}=i\pi_\varphi\,,
\end{equation}
and anti-BRST transformations
\begin{equation}
\bar{\delta}_1\pi^a=\bar{C}\phi^a\,,\,\,\,\,\,\bar{\delta}_1\varphi=-\dot{\bar{C}}\,,\,\,\,\,\,\bar{\delta}_1C=i\frac{\phi^a\phi^b\phi^a\phi^b}{f\phi^c\pi^c}+\bar{C}\frac{\phi^a\phi^b}{\phi^c\phi^c}C
\,,
\end{equation}
\begin{equation}
\bar{\delta}_2\pi^a=-\bar{C}\phi^a\,,\,\,\,\,\,\bar{\delta}_2\varphi=-\dot{\bar{C}}\,,\,\,\,\,\,\bar{\delta_2}{C}=i\pi_\varphi\,.
\end{equation}
This ends our analysis of the $O(N)$ nonlinear sigma model as a particular example of the generalized rigid rotor.

\section{Conclusion and Final Remarks}
We have identified a family of similar second-class dynamical systems encompassing motion along conic curves \cite{Barbosa:2018dmb}, the rigid rotor \cite{Nemeschansky:1987xb}, and the nonlinear sigma model as particular cases, which can be cast into first-class exhibiting gauge symmetry.   The main feature of these similar systems consists of a second-class geometrical constraint which is directly implemented in the action via a Lagrange multiplier.  One of the simplest cases is exactly the rigid rotor, which we have discussed in section {\bf 5} -- in that case the constraint hypersurface is a $S^n$ hypersphere in $n$ dimensions.  The generalization of that hypersphere to a more arbitrary smooth geometric hypersurface $T(q^k)=0$ lead to our general system (\ref{H}) which we named as generalized rigid rotor.   We have shown how these systems can be easily made gauge invariant and pursued their corresponding BFV functional quantization.  A general expression for the BRST charge was also explicitly given.  In this way, it was possible to investigate the common properties of such systems, particularly those related to BRST symmetries, in a unified way.  It is worth noting that, in contrast to other abelianization methods, no auxiliary variables were needed.   For the gauge invariant general model -- c.f. action (\ref{S}) with Hamiltonian (\ref{H}), invariant under (\ref{ginv}) -- we have made use restrictly of the already existing variables in phase space.  Whilst for obtaining the BRST symmetry, we have introduced only the proper prescribed Grassmann variables concerning the ghost fields.   The examples have also shown that in certain cases it is much easier and clear to handle the formulae along the general dynamical system, with the familiar compact index notations and only afterwards turn to the specific notation for the desired particular systems.   Particularities of the gauge choice become also more transparent through the proposed general perspective.  The abelianization of some of these models has been studied in the literature, always on isolated terms, through different methods -- for instance the BFFT conversion method has been recurrently applied to the nonlinear sigma model \cite{Abreu:2000ip, Hong:1999gx, Hong:2002vr, BarcelosNeto:1997kn}.  We hope that the general approach presented here may consolidate further treatments of these similar systems in a unified way.

\end{document}